\documentstyle[aps,epsf]{revtex}  

\begin{document}        

\def\be{\begin{eqnarray}}
\def\en{\end{eqnarray}}
\def\non{\nonumber}
\def\la{\langle}
\def\ra{\rangle}
\def\nc{N_c^{\rm eff}}
\def\vp{\varepsilon}
\def\vma{{_{V-A}}}
\def\vpa{{_{V+A}}}
\def\m{\hat{m}}
\def\lsim{ {\ \lower-1.2pt\vbox{\hbox{\rlap{$<$}\lower5pt\vbox{\hbox{$\sim$}
}}}\ } }
\def\ov{\overline}
\baselineskip 14pt
\title{Charmless Hadronic Two-body Decays of $B$ Mesons}
\author{Hai-Yang Cheng}
\address{Institute of Physics, Academia Sinica,
Taipei, Taiwan 115, Republic of China}
%
\maketitle              

\begin{abstract}        
A way of circumventing the gauge and infrared problems with
effective Wilson coefficients is shown. Implications of
experimentally measured charmless $B$ decays are discussed.
\
\end{abstract}      

\section{Introduction}               

In past years we have witnessed remarkable progress in the study
of exclusive charmless $B$ decays. Experimentally, CLEO
\cite{CLEO} has discovered many new two-body decay modes
$B\to\eta' K^\pm,~\eta' K^0,~\pi^\pm K^0,~\pi^\pm K^\mp,~\pi^0
K^\pm,~\rho^0\pi^\pm,~\omega K^\pm$ and found a possible evidence
for $B\to \phi K^*$. Moreover, CLEO has provided new improved
upper limits for many other decay modes. While all the measured
channels are penguin dominated, the most recently observed
$\rho^0\pi^-$ mode is dominated by the tree diagram. In the
meantime, updates and new results of many $B\to PV$ decays with
$P=\eta,\eta',\pi,K$ and $V=\omega,\phi,\rho, K^*$ as well as
$B\to PP$ decays will be available soon. With the $B$ factories
Babar and Belle starting to collect data, many exciting and
harvest years in the arena of $B$ physics and $CP$ violation are
expected to come. Theoretically, many significant improvements and
developments have been achieved over past years. For example, a
next-to-leading order effective Hamiltonian for current-current
operators and QCD as well as electroweak penguin operators becomes
available. The renormalization scheme and scale problems with the
factorization approach for matrix elements can be circumvented by
employing scale- and scheme-independent effective Wilson
coefficients. Heavy-to-light form factors have been computed using
QCD sum rules, lattice QCD and potential models.

\section{Effective Wilson Coefficients}
Although the hadronic matrix element $\la O(\mu)\ra$ can be
directly calculated in the lattice framework, it is conventionally
evaluated under the factorization hypothesis so that $\la
O(\mu)\ra$ is factorized into the product of two matrix elements
of single currents, governed by decay constants and form factors.
In spite of its tremendous simplicity, the naive factorization
approach encounters two principal difficulties.  One of them is
that the hadronic matrix element under factorization is
renormalization scale $\mu$ independent as the vector or
axial-vector current is partially conserved. Consequently, the
amplitude $c_i(\mu) \la O\ra_{\rm fact}$ is not truly physical as
the scale dependence of Wilson coefficients does not get
compensation from the matrix elements. A plausible solution to the
aforementioned scale problem is to extract the $\mu$ dependence
from the matrix element $\la O(\mu)\ra$, combine it with the
$\mu$-dependent Wilson coefficients to form $\mu$-independent
effective Wilson coefficients.  However, it was pointed out
recently in \cite{Buras98} that $c_i^{\rm eff}$ suffer from the
gauge and infrared ambiguities since an off-shell external quark
momentum, which is usually chosen to regulate the infrared
divergence occurred in the radiative corrections to the local
4-quark operators, will introduce a gauge dependence. Therefore,
this solution, though removes the scale and scheme dependence of a
physical amplitude in the framework of the factorization
hypothesis, often introduces the infrared cutoff and gauge
dependence.

It was recently shown in \cite{CLY} that the above-mentioned
problems on gauge dependence and infrared singularity associated
with the effective Wilson coefficients can be resolved by
perturbative QCD (PQCD) factorization theorem. In this formalism,
partons, {\it i.e.}, external quarks, are assumed to be on shell,
and both ultraviolet and infrared divergences in radiative
corrections are isolated using the dimensional regularization.
Because external quarks are on shell, gauge invariance of the
decay amplitude is maintained under radiative corrections to all
orders. This statement is confirmed by an explicit one-loop
calculation in \cite{CLY}. The obtained ultraviolet poles are
subtracted in a renormalization scheme, while the infrared poles
are absorbed into universal nonperturbative bound-state wave
functions. The remaining finite piece is grouped into a hard decay
subamplitude. The decay rate is then factorized into the
convolution of the hard subamplitude with the bound-state wave
functions, both of which are well-defined and gauge invariant.
Explicitly, the effective Wilson coefficient has the expression
\begin{eqnarray}
c^{\rm eff}=c(\mu)g_1(\mu)g_2(\mu_f)\;, \label{nef}
\end{eqnarray}
where $g_1(\mu)$ is an evolution factor from the scale $\mu$ to
$m_b$, whose anomalous dimension is the same as that of $c(\mu)$,
and $g_2(\mu_f)$ describes the evolution from $m_b$ to $\mu_f$
($\mu_f$ being a factorization scale arising from the dimensional
regularization of infrared divergences), whose anomalous dimension
differs from that of $c(\mu)$ because of the inclusion of the
dynamics associated with spectator quarks. Setting $\mu_f=M_b$,
the effective Wilson coefficients obtained from the one-loop
vertex corrections to the 4-quark operators $O_i$ have the form:
\be
\label{ceff} c_i^{\rm eff}\Big|_{\mu_f=m_b} &=&
c_i(\mu)+{\alpha_s\over 4\pi}\left(\gamma^{(0)T}\ln{m_b\over
\mu}+\hat r^T\right)_{ij}c_j(\mu)+\cdots,
\en
where the matrix $\hat r$ gives momentum-independent constant
terms which depend on the treatment of $\gamma_5$. The expression
of $\hat r$ for $\Delta B=1$ transition current-current operators
$O_1,O_2$ is given in \cite{CLY}, while the complete result for
QCD-penguin operators $O_3,\cdots,O_6$ and electroweak penguin
operators $O_7,\cdots,O_{10}$ is given in \cite{CCTY}. It should
be accentuated that, contrary to the previous work based on Landau
gauge and off-shell regularization \cite{Ali}, the matrix $\hat r$
given in Eq. (\ref{ceff}) is gauge invariant. Consequently, the
effective Wilson coefficients (\ref{ceff}) are not only scheme and
scale independent but also free of gauge and infrared problems.

\section{Effective parameters and nonfactorizable effects}

It is known that the effective Wilson coefficients appear in the
factorizable decay amplitudes in the combinations $a_{2i}=
{c}_{2i}^{\rm eff}+{1\over N_c}{c}_{2i-1}^{\rm eff}$ and
$a_{2i-1}= {c}_{2i-1}^{\rm eff}+{1\over N_c}{c}^{\rm eff}_{2i}$
$(i=1,\cdots,5)$. Phenomenologically, the number of colors $N_c$
is often treated as a free parameter to model the nonfactorizable
contribution to hadronic matrix elements and its value can be
extracted from the data of two-body nonleptonic decays. As shown
in \cite{Cheng94}, nonfactorizable effects in the decay amplitudes
of $B\to PP,~VP$ can be absorbed into the parameters $a_i^{\rm
eff}$. This amounts to replacing $N_c$ in $a^{\rm eff}_i$ by
$(N_c^{\rm eff})_i$. Explicitly,
\be
a_{2i}^{\rm eff}={c}_{2i}^{\rm eff}+{1\over (N_c^{\rm
eff})_{2i}}{c}_{2i-1}^{ \rm eff}, \qquad \quad a_{2i-1}^{\rm eff}=
{c}_{2i-1}^{\rm eff}+{1\over (N_c^{\rm eff})_{2i-1}}{c}^{\rm
eff}_{2i}, \qquad (i=1,\cdots,5),
\en
where $(1/N_c^{\rm eff})_i\equiv (1/N_c)+\chi_i$ with $\chi_i$
being the nonfactorizable terms which receive main contributions
from color-octet current operators \cite{Neubert}. In the absence
of final-state interactions, we shall assume that $\chi_i$ and
hence $\nc$ are real. If $\chi_i$ are universal (i.e. process
independent) in charm or bottom decays, then we have a generalized
factorization scheme in which the decay amplitude is expressed in
terms of factorizable contributions multiplied by the universal
effective parameters $a_i^{\rm eff}$. Phenomenological analyses of
the two-body decay data of $D$ and $B$ mesons indicate that while
the generalized factorization hypothesis in general works
reasonably well, the effective parameters $a_{1,2}^{\rm eff}$ do
show some variation from channel to channel, especially for the
weak decays of charmed mesons. A recent updated analysis of $B\to
D\pi$ data gives \cite{CY} \be \label{a2} \nc(B\to D\pi)\sim
(1.8-2.2),\qquad\quad \chi_2(B\to D\pi)\sim (0.12-0.21).
\en

It is customary to assume in the literature that $(N_c^{\rm
eff})_1 \approx (N_c^{\rm eff})_2\cdots\approx (N_c^{\rm
eff})_{10}$ so that the subscript $i$ can be dropped; that is, the
nonfactorizable term is usually postulated to behave in the same
way in penguin and tree decay amplitudes. A closer investigation
shows that this is not the case. It has been argued in \cite{CT98}
that nonfactorizable effects in the matrix elements of
$(V-A)(V+A)$ operators are {\it a priori} different from that of
$(V-A)(V-A)$ operators. One primary reason is that the Fierz
transformation of the $(V-A)(V+A)$ operators $O_{5,6,7,8}$ is
quite different from that of $(V-A)(V-A)$ operators $O_{1,2,3,4}$
and $O_{9,10}$. As a result, contrary to the common assertion,
$\nc(LR)$ induced by the $(V-A)(V+A)$ operators are theoretically
different from $\nc(LL)$ generated by the $(V-A)(V-A)$ operators
\cite{CT98}. Therefore, we shall assume that
\be
&& N_c^{\rm eff}(LL)\equiv \left(N_c^{\rm
eff}\right)_1\approx\left(N_c^{\rm eff}\right)_2\approx
\left(N_c^{\rm eff}\right)_3\approx\left(N_c^{\rm
eff}\right)_4\approx \left(N_c^{\rm eff}\right)_9\approx
\left(N_c^{\rm eff}\right)_{10},   \non\\ && N_c^{\rm
eff}(LR)\equiv \left(N_c^{\rm eff}\right)_5\approx\left(N_c^{\rm
eff}\right)_6\approx \left(N_c^{\rm eff}\right)_7\approx
\left(N_c^{\rm eff}\right)_8,
\en
and $N_c^{\rm eff}(LR)\neq N_c^{\rm eff}(LL)$ in general. Since
the energy release in the energetic two-body charmless $B$ decays
is not less than that in $B\to D\pi$ decays, it is thus expected
that $ |\chi({\rm two-body~rare~B~decay})|\lsim |\chi(B\to D\pi)|
$ and hence $\nc(LL)\approx \nc(B\to D\pi)\sim 2$. From the data
analysis in the next section, we shall see that $\nc(LL)<3$ and
$\nc(LR)>3$.

\section{Analysis of Data}
We have studied in detail the two-body charmless $B$ decays for
$B_{u,d}$ mesons in \cite{CCTY} and for $B_s$ mesons in
\cite{CCT}. In what follows we show some highlights of the data
analysis.
\subsection{Spectator-dominated rare $B$ decays}
Very recently, CLEO has made the first observation of a hadronic
$b\to u$ decay, namely $B^\pm\to\rho^0\pi^\pm$ \cite{Gao,Frank}.
The preliminary measurement yields:
\be
{\cal B}(B^\pm\to\rho^0\pi^\pm)=\,(1.5\pm 0.5\pm 0.4)\times
10^{-5}.
\en
The branching ratios of this mode decreases with $\nc(LL)$ as it
involves interference between external and internal $W$-emission
amplitudes. From Fig. 1  it is clear that this class-III decay is
sensitive to $1/\nc$ if $\nc(LL)$ is treated as a free parameter,
namely, $\nc(LR)=\nc(LL)=\nc$; it has the lowest value of order
$1\times 10^{-6}$ and then grows with $1/\nc$. We see from Fig.~1
that $0.35\leq 1/\nc\leq 0.92$. Since the tree diagrams make the
dominant contributions, we then have
\be
1.1\leq \nc(LL)\leq 2.6~~~{\rm from}~B^\pm\to\rho^0\pi^\pm.
\en
Therefore, $\nc(LL)$ is favored to be less than 3, as expected.

\begin{figure}[ht]
\centerline{\epsfxsize 2.2 truein \epsfbox{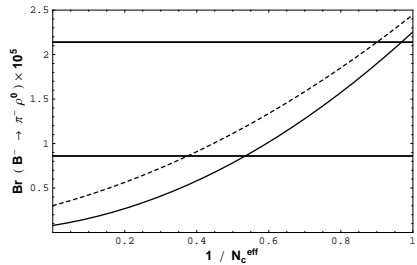}}
\vskip .2cm
    \caption[]{\small The branching ratio of $B^-\to\rho^0\pi^-$ versus
    $1/\nc$. The solid (dotted) curve is calculated using the
    Baure-Stech-Wirbel (light-cone sum rule)
     model for form factors, while the solid thick lines are the CLEO measurements
     with one sigma errors.}
\end{figure}

There are two additional experimental hints that favor the choice
$\nc(LL)\sim 2$. First is the class-III decay
$B^\pm\to\pi^\pm\omega$. This mode is very similar to
$\rho^0\pi^\pm$ as its decay amplitude differs from that of
$\omega\pi^\pm$ only in the $2(a_3+a_5)$ penguin term which is
absent in the former. Since the coefficient $(a_3+a_5)$ is small
and it is further subject to the quark-mixing angle suppression,
the decay rates of $\omega\pi^\pm$ and $\rho^0\pi^\pm$ are very
similar. Although experimentally only the upper limit ${\cal
B}(B^\pm\to\pi^\pm\omega)<2.3\times 10^{-5}$ is quoted by CLEO
\cite{CLEOomega2}, the CLEO measurements ${\cal B}(B^\pm\to
K^\pm\omega)=(1.5^{+0.7}_{-0.6}\pm 0.2) \times 10^{-5}$ and ${\cal
B}(B^\pm\to h^\pm\omega)=(2.5^{+0.8}_{-0.7}\pm 0.3)\times 10^{-5}$
with $h=\pi,~K$ indicate that the central value of ${\cal
B}(B^\pm\to\pi^\pm\omega)$ is about $1\times 10^{-5}$. This means
$0.4<1/\nc(LL)<0.6$ (see Fig.~3 of \cite{CCTY}) or
$1.7<\nc(LL)<2.5$ is favored; the prediction for $\nc(LL)=2$ is
${\cal B}(B^\pm\to\omega\pi^\pm)= 0.8\times 10^{-5}$ and
$1.1\times 10^{-5}$ in the Bauer-Stech-Wirbel (BSW) model
\cite{BSW85} and the light-cone sum rule (LCSR) analysis
\cite{Ball} for heavy-to-light form factors, respectively. The
second hint comes from the decay $B^0\to\pi^+\pi^-$. A very recent
CLEO analysis of $B^0\to\pi^+\pi^-$ presents an improved upper
limit, ${\cal B}(B^0\to\pi^+\pi^-)<0.84\times 10^{-5}$ \cite{Roy}.
It also implies that $\nc(LL)$ is preferred to be smaller
\cite{CCTY}.

\subsection{$B\to \phi K,~\phi K^*$ decays}
The decay amplitudes of the penguin-dominated modes $B\to \phi K$
and $B\to \phi K^*$ are governed by the effective coefficients
$[a_3+a_4+a_5-{1\over 2}(a_7+a_9+a_{10})]$. Note that the QCD
penguin coefficients $a_3$ and $a_5$ are sensitive to $\nc(LL)$
and $\nc(LR)$, respectively. We see from Figs.~6 and 7 of
\cite{CCTY} that the decay rates of $B\to \phi K^{(*)}$ increase
with $1/\nc(LR)$ irrespective of $\nc(LL)$. The new CLEO upper
limit ${\cal B}(B^\pm\to\phi K^\pm)< 0.59\times 10^{-5}$ at 90\%
C.L. \cite{Frank} implies that \be \label{ncLR} \nc(LR)\geq
\cases{4.2 & BSW, \cr 3.2 & LCSR,  \cr}
\en
with $\nc(LL)$ being fixed at the value of 2. Note that this
constraint is subject to the corrections from spacelike penguin
and $W$-annihilation contributions. At any rate, it is safe to
conclude that $\nc(LR)>3>\nc(LL)$.

CLEO has seen a $3\sigma$ evidence for the decay $B\to\phi K^*$.
Its branching ratio, the average of $\phi K^{*-}$ and $\phi
K^{*0}$ modes, is reported to be \cite{Frank}
\be
\label{phiK*} {\cal B}(B\to\phi K^*)\equiv {1\over 2}\left[{\cal
B}(B^\pm\to\phi K^{*\pm}) +{\cal B}(B^0\to\phi K^{*0})\right]
=\left(1.1^{+0.6}_{-0.5}\pm 0.2\right)\times 10^{-5}.
\en
We find that the branching ratio of $B\to\phi K^*$ is in general
larger (less) than that of $B\to\phi K$ in the LCSR (BSW) model.
This is because $\Gamma(B\to \phi K^*)$ is very sensitive to the
form factor ratio $x=A_2^{BK^*}(m^2_\phi)/ A_1^{BK^*}(m^2_\phi)$,
which is equal to 0.875 (1.03) in the LCSR (BSW) model. In
particular, ${\cal B}(B\to\phi K^*)=0.77\times 10^{-5}$ is
predicted by the LCSR for $\nc(LL)=2$ and $\nc(LR)=5$, which is in
accordance with experiment. It is evident that the data of
$B\to\phi K$ and $B\to\phi K^*$ can be simultaneously accommodated
in the LCSR analysis (see Figs.~6 and 7 of \cite{CCTY}).
Therefore, the non-observation of $B\to\phi K$ does not
necessarily invalidate the factorization hypothesis; it could
imply that the form-factor ratio $A_2/A_1$ is less than unity. Of
course, it is also possible that the absence of $B\to\phi K$
events is a downward fluctuation of the experimental signal. At
any rate, in order to clarify this issue and to pin down the
effective number of colors $\nc(LR)$, measurements of $B\to\phi K$
and $B\to\phi K^*$ are urgently needed with sufficient accuracy.

\subsection{$B\to\eta' K$ and $\eta K$ decays}
The published CLEO results \cite{Behrens1} on the decay $B\to\eta'
K$: ${\cal B}(B^\pm\to\eta' K^\pm) = \left(6.5^{+1.5}_{-1.4}\pm
0.9\right)\times 10^{-5}$ are several times larger than earlier
theoretical predictions \cite{Chau1,Kramer,Du} in the range of
$(1-2)\times 10^{-5}$. It was pointed out in past two years by
several authors \cite{Ali,Kagan} that the decay rate of $B\to\eta'
K$ will get enhanced because of the small running strange quark
mass at the scale $m_b$ and sizable $SU(3)$ breaking in the decay
constants $f_8$ and $f_0$.

As shown in \cite{CCTY}, if $\nc(LL)$ is treated to be the same as
$\nc(LR)$, the branching ratio of $(B^-\to\eta' K^-)$ is $\sim
(2.7-4.7)\times 10^{-5}$ at $0<1/\nc<0.5$ and it becomes
$(3.5-3.8)\times 10^{-5}$ when the $\eta'$ charm content
contribution with $f_{\eta'}^c=-6.3$ MeV is taken into account.
However, the discrepancy between theory and experiment is greatly
improved by treating $\nc(LL)$ and $\nc(LR)$ differently. Setting
$\nc(LL)=2$, we find that (see Fig.~2) the decay rates of
$B\to\eta' K$ are considerably enhanced especially at small
$1/\nc(LR)$. Specifically, ${\cal B}(B^\pm\to\eta' K^\pm)$ at
$1/\nc(LR)\leq 0.2$ is enhanced from $(3.6-3.8)\times 10^{-5}$ to
$(4.6-6.1) \times 10^{-5}$ due to three enhancements. First, the
$\eta'$ charm content contribution $a_2X_c^{(BK,\eta')}$ now
always contributes in the right direction to the decay rate
irrespective of the value of $\nc(LR)$. Second, the interference
in the spectator amplitudes of $B^\pm\to\eta' K^\pm$ is
constructive. Third, the term proportional to
$2(a_3-a_5)X_u^{(BK,\eta')}+(a_3+a_4-a_5)X_s^{(BK,\eta')}$ is
enhanced when $(\nc)_3=(\nc)_4=2$. Therefore, the data of $B\to
K\eta'$ provide a strong support for $\nc(LL)\sim 2$ and the
relation $\nc(LR)>\nc(LL)$. The mode $B\to\eta' K$ has the largest
branching ratio in the two-body charmless $B$ decays due mainly to
the constructive interference between the penguin contributions
arising from the $(\bar uu+\bar dd)$ and $\bar ss$ components of
the $\eta'$. By contrast, the destructive interference for the
$\eta$ production leads to a much small decay rate for $B\to\eta
K$.

\begin{figure}[tb]
\centerline{\epsfxsize 2.2 truein \epsfbox{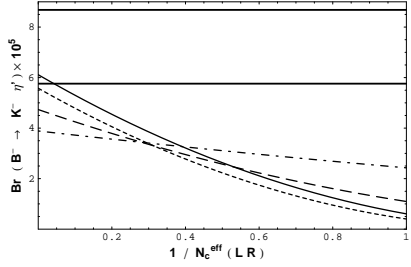}}
\vspace{0.2cm}
    \caption[]{\small The branching ratio of $B^\pm\to\eta' K^\pm$ as a
   function of $1/\nc(LR)$ with $\nc(LL)$ being fixed at the value of 2
   and $\eta=0.370$, $\rho=0.175$, $m_s(m_b)=90$ MeV. The
   calculation is done using the BSW model for form factors.
   The charm content of the $\eta'$ with $f_{\eta'}^c=-6.3\,{\rm MeV}$
   contributes to the solid curve but not to the dotted curve.
   The anomaly contribution to $\la\eta'|\bar s\gamma_5s|0\ra$
   is included. For comparison, predictions for
   $\nc(LL)=\nc(LR)$ as depicted by the dashed curve with $f_{\eta'}^c
   =0$ and dot-dashed curve with $f_{\eta'}^c=-6.3$ MeV are also shown.
    The solid thick lines are the preliminary updated
    CLEO measurements \cite{Gao,Frank} with one sigma errors.}
   \label{fig:Ketap}
\end{figure}

\subsection{$B^\pm\to \omega K^\pm$ and $B^\pm\to\rho^0 K^\pm$ decays}
  The CLEO observation \cite{CLEOomega2} of a large branching ratio for
$B^\pm\to \omega K^\pm$, $ {\cal B}(B^\pm\to\omega
K^\pm)=\left(1.5^{+0.7}_{-0.6}\pm 0.2\right) \times 10^{-5}$, is
rather difficult to explain at first sight.  We showed in
\cite{CCTY} that in the absence of FSI, the branching ratio of
$B^+\to\omega K^+$ is expected to be of the same order as ${\cal
B}(B^+\to\rho^0 K^+)\sim (0.5-1.0)\times 10^{-6}$, while
experimentally it is of order $1.5\times 10^{-5}$. We argued that
$B^+\to\omega K^+$ receives a sizable final-state rescattering
contribution from the intermediate states
$K^{*-}\pi^0,K^{*-}\eta,K^{*0}\pi^-,K^-\rho^0,K^0\rho^-$ which
interfere constructively, whereas the analogous rescattering
effect on $B^+\to\rho^0 K^+$ is very suppressed. However, if the
measured branching ratio $\rho^0 K^+$ is similar to $\omega K^+$,
then $W$-annihilation and spacelike penguins may play a prominent
role. Likewise, the decay mode $B^0\to K^+K^-$ is expected to be
dominated by inelastic rescattering from $\rho^+\rho^-,\pi^+\pi^-$
intermediate states.



\begin{references}  
\newcommand{\bi}{\bibitem}

\bi{CLEO} For a review of CLEO measurements on charmless $B$
decays, see K. Lingel, T. Skwarnicki, and J.G. Smith,
hep-ex/9804015.

\bi{Buras98} A.J. Buras and L. Silvestrini, hep-ph/9806278.

\bi{CLY} H.Y. Cheng, H.n. Li, and K.C. Yang,  hep-ph/9902239.

\bi{CCTY} Y.H. Chen, H.Y. Cheng, B. Tseng, and K.C. Yang,
IP-ASTP-01-99 (1999).

\bibitem {Ali} A. Ali and C. Greub, Phys. Rev. D {\bf 57}, 2996 (1998).

\bi{Cheng94} H.Y. Cheng, Phys. Lett. B {\bf 395}, 345 (1994); A.N.
Kamal and A.B. Santra, Alberta Thy-31-94 (1994); Z. Phys. C {\bf
72}, 91 (1996); J.M. Soares, Phys. Rev. D {\bf 51}, 3518 (1995).

\bi{Neubert} M. Neubert and B. Stech, in {\it Heavy Flavours},
edited by A.J. Buras and M. Lindner, 2nd ed. (World Scientific,
Singapore, 1998), p.294.

\bi{CY} H.Y. Cheng and K.C. Yang, hep-ph/9811249, to appear in
Phys. Rev. D.

\bi{CT98} H.Y. Cheng and B. Tseng, Phys. Rev. D {\bf 58}, 094005
(1998).

\bi{CCT} Y.H. Chen, H.Y. Cheng, and B. Tseng, Phys. Rev. D {\bf
59}, 074003 (1999).

\bi{Gao} CLEO Collaboration, Y.S. Gao, in this proceedings.

\bi{Frank}CLEO Collaboration, F. W\"urthwein, in this proceedings.

\bi{CLEOomega2} CLEO Collaboration, T. Bergfeld {\it et al.,}
Phys. Rev. Lett. {\bf 81}, 272 (1998).

\bibitem{BSW85} M. Wirbel, B. Stech, and M. Bauer, Z. Phys. C {\bf 29},
637 (1985).

\bi{Ball} P. Ball and V.M. Braun, Phys. Rev. D {\bf 58}, 094016
(1998); P. Ball,  J. High Energy Phys. {\bf 9809}, 005 (1998).

\bi{Roy} CLEO Collaboration, J. Roy, invited talk presented at the
XXIX International Conference on High Energy Physics, Vancouver,
July 23-28, 1998.

\bi{Behrens1} CLEO Collaboration, B.H. Behrens {\it et al.,} Phys.
Rev. Lett. {\bf 80}, 3710 (1998).

\bi{Chau1} L.L. Chau, H.Y. Cheng, W.K. Sze, H. Yao, and B. Tseng,
Phys. Rev. D {\bf 43}, 2176 (1991); {\bf 58}, (E)019902 (1998).

\bi{Kramer} G. Kramer, W.F. Palmer, and H. Simma, Z. Phys. C {\bf
66}, 429 (1995); Nucl. Phys. B {\bf 428}, 77 (1994).

\bi{Du} D.S. Du and L. Guo, Z. Phys. C {\bf 75}, 9 (1997).

\bi{Kagan} A.L. Kagan and A.A. Petrov, hep-ph/9707354; N.G.
Deshpande, B. Dutta, and S. Oh, Phys. Rev. D {\bf 57}, 5723
(1998).

 \end{references}
\end{document}